\pdfoutput=1
\documentclass[a4paper,traditabstract]{aa}   
\usepackage{graphicx}
\usepackage{amsmath}
\usepackage{natbib}
\usepackage{comment}
\usepackage{txfonts}
\usepackage{bm}
\usepackage[dvipsnames]{xcolor}
\usepackage[breaklinks=true]{hyperref}
\hypersetup{
  colorlinks   = true,
  urlcolor     = blue,
  linkcolor    = blue,
  citecolor    = blue,
  breaklinks   = true
}
\bibpunct{(}{)}{;}{a}{}{,}
\begin{document}

\title{Two-dimensional weak lensing shear for cluster mass, concentration, ellipticity and miscentering estimation}
\titlerunning{Two-dimensional weak lensing shear for cluster mass estimation}

\author{C. Murray\inst{1} \and M. Kilbinger\inst{1} \and C. Payerne \inst{2}
        }

\institute{
Université Paris-Saclay, Université Paris Cité, CEA, CNRS, AIM, 91191 Gif-sur-Yvette, France
\and
Université Paris-Saclay, CEA, IRFU, 91191 Gif-sur-Yvette, France}

\date{Received [date] / Accepted [date]}

\abstract{
The effects of weak gravitational lensing induced by clusters of galaxies makes it possible to constrain their dark matter distribution. In general mass estimates from weak lensing are made by compressing the two-dimensional shear-field into one-dimension. In this work we build a forward model of the two-dimensional reduced-shear field around a galaxy cluster, including the non-Gaussian correlated large-scale structure from N-body simulations, the uncorrelated line-of-sight structure, cluster miscentering and ellipticity. We then use simulation-based inference to obtain posteriors on the cluster mass, concentration, miscentering and ellipticity from the full field. The two-dimensional approach leads to improvements on cluster mass constraints by up to 20\%. We show that this is due to a better treatment of the spatially correlated noise across the shear field. There is also significant improvements on the cluster concentration and miscentering. Whilst improvements are also made to the estimation of cluster ellipticity, the constraints are still prior dependent for these individual cluster constraints. We also quantify the impact of the line-of-sight structure on the constraints. Neglecting it underestimates the mass uncertainty by up to a factor of $2$ for the most massive clusters. 
}


\maketitle

\section{Introduction}
\label{sec:introduction}

As light from distant galaxies propagates through the Universe, its path is perturbed by the gravitational field of massive objects, such as galaxy clusters. This phenomenon is referred to as gravitational lensing \citep{1992Schneider} and leads to many observable effects, including the deformation of the images of distant galaxies. In the strong-lensing regime this produces dramatic lensing arcs, while in the regime we study here, weak gravitational lensing, it produces a subtle shear, deflection and magnification of galaxy images \citep{Bartelmann2001}.

As the magnitude of these effects depends upon the mass of the galaxy cluster, we can use observations of gravitational lensing to estimate galaxy cluster masses \citep[for example][]{Lesci2025,Murray2022,McClintock2019,Okabe2016}. An advantage of gravitational lensing over other mass estimation methods is that it does not depend on the dynamical state of the galaxy cluster. Strong gravitational lensing can provide strong constraints on the matter distribution in the centre of clusters, but relies on fortuitous alignments between the observer, the cluster, and a background galaxy. Therefore such observations are not possible for every cluster. Conversely, weak gravitational lensing does not rely on these chance alignments, and we are therefore able to constrain the mass of any cluster using weak lensing, albeit with large errors. In this work, we study specifically weak gravitational shear. Cluster mass constraints can also be made from magnification \citep[for example][]{Murray2025a,Chiu2020,Tudorica2017}, however in general shear can provide stronger constraints, essentially because the variance in galaxy shapes is smaller than the variance in galaxy magnitudes, galaxy sizes and their density across the sky (which is what magnification is sensitive to).

In this work we develop a method which uses the two-dimensional shear field around galaxy clusters to measure the cluster mass, concentration, ellipticity and the miscentering of the total mass distribution compared to the identified cluster position. 

Using the two-dimensional shear field, rather than the azimuthally averaged, one-dimensional profile of the tangential shear around clusters, has two advantages. Firstly, both miscentering and ellipticity of the cluster mass distribution create non-circular signals in the shear field, which are lost to the circular average. The two-dimensional field allows us to directly measure these signals. Secondly, the cosmic web will itself shear the images of galaxies, adding a source of noise for cluster mass estimation which is correlated across the cluster field \citep{Hoekstra2003, Wu2019, Murray2025a, Vecchi2025}. When azimuthally averaging, this correlated noise is combined in a sub-optimal manner \citep{Murray2025a}. In Section~\ref{subsec:clss_results} we measure the size of both effects, by comparing forward models with and without the line-of-sight structure.

In this work we forward model the entire signal, including the full non-Gaussian signal from the cosmic web. The result of building this model makes it possible to also measure cluster miscentering and ellipticity. Previous measurements of halo ellipticity from two-dimensional weak-lensing shear \citep[e.g.][]{Robison2023,Schrabback2021,Dvornik2019,Oguri2010} have not fully included the pixel-to-pixel covariance from large-scale structure. In our forward model this covariance is included and in a non-Gaussian manner.

In Section \ref{sec:theory} we describe weak lensing around galaxy clusters. In Section \ref{sec:simulations} we discuss the details of our simulations and the construction of our one-dimensional and two-dimensional summary statistics. In Section \ref{sec:results} we present an analysis of the different noise contributions to cluster shear profiles and a comparison between the one-dimensional and two-dimensional results for the estimation of the cluster parameters. In Section \ref{sec:discussion} we discuss the results and the applicability of these methods to cosmological surveys of galaxy clusters. 

\section{Weak gravitational lensing around clusters}
\label{sec:theory}

The effects of lensing by clusters can be conceptually understood within the framework of the halo model \citep{cooray2002}, in which the large-scale structure of the Universe is separated into two contributions: the one-halo term and the two-halo term. The one-halo term is the entire mass of the cluster (principally dark matter, but also galaxies, stars and gas). It is from this component that we estimate the cluster mass, concentration, ellipticity and miscentering. The two-halo term is the correlated large-scale structure surrounding the galaxy cluster. As we will see, it is this component that is the most challenging to model for this work. 

In this section we summarise the lensing quantities needed to build the forward model and refer the reader to \citet{Bartelmann2001} and \citet{Umetsu2020} for comprehensive reviews of weak lensing and cluster lensing respectively. Throughout, the shear $\gamma = \gamma_1 + {\rm i}\gamma_2$ is a spin-2 field: it transforms as $\gamma \to \gamma\,{\rm e}^{2{\rm i}\varphi}$ under a rotation of the coordinate frame by $\varphi$. 

Weak gravitational lensing deforms the observed shapes of galaxies such that, 

\begin{equation}
    \epsilon_{\rm obs} = \epsilon_{\rm int} + \frac{\gamma}{1-\kappa},
    \label{eq:ellipticity_obs}
\end{equation}

where $\epsilon_{\rm int}$ is the intrinsic ellipticity of the galaxy, $\gamma$ is the weak lensing shear and $\kappa$ is the convergence. We can therefore define the reduced shear,

\begin{equation}
    g = \frac{\gamma}{1-\kappa} .
    \label{eq:reduced_shear}
\end{equation}

Both the shear and the convergence are determined by the mass distribution of the lens. The convergence is the projected surface mass density $\Sigma$ divided by the critical surface mass density,

\begin{equation}
    \kappa(\vec{\theta}) = \frac{\Sigma(\vec{\theta})}{\Sigma_{\rm crit}},
    \qquad
    \Sigma_{\rm crit} = \frac{c^2}{4 \pi G} \frac{D_{\rm s}}{D_{\rm d}\, D_{\rm ds}},
    \label{eq:convergence}
\end{equation}

where $D_{\rm s}$, $D_{\rm d}$ and $D_{\rm ds}$ are the angular diameter distances to the source, to the lens, and between the lens and the source. We evaluate $\Sigma_{\rm crit}$ at the mean redshift of the source distribution (Sect.~\ref{subsec:ulss}).

Around a cluster it is useful to separate the shear into a tangential component $\gamma_{\rm t}$ and a cross component $\gamma_{\times}$, defined with respect to the line connecting each position to the cluster centre. For a circularly symmetric mass distribution the cross component vanishes, and the tangential shear is directly related to the excess surface mass density,
\begin{equation}
    \gamma_{\rm t}(R) = \frac{\Delta\Sigma(R)}{\Sigma_{\rm crit}},
    \qquad
    \Delta\Sigma(R) = \bar{\Sigma}(<R) - \Sigma(R),
    \label{eq:esd}
\end{equation}
where $\bar{\Sigma}(<R)$ is the mean surface mass density within the projected radius $R$.

\subsection{One-halo term}

For the one-halo term we use an elliptical modification of the Navarro-Frenk-White (NFW) profile \citep{Navarro1997}. The standard NFW profile is characterized by two parameters: the mass enclosed within the radius at which the mean enclosed density is 200 times the mean matter density of the Universe, $M_{200m}$, and the concentration $c_{200m}$. The shear and convergence around spherical NFW halos has been presented in \citet{Wright2000}. We incorporate ellipticity into the shear and convergence fields following \citet{Adhikari2015} (see also \citealt{Natarajan2000, Mandelbaum2006, Clampitt2016}).

The tangential and cross components of the excess surface mass density of an elliptical NFW halo are written to first order in the ellipticity $e$, as a monopole plus a quadrupole,
\begin{align}
    \Delta\Sigma_{\rm t}(R, \psi) &= \Delta\Sigma_{0}(R) + \frac{e}{2}\, Q_{+}(R) \cos 2\psi ,
    \label{eq:esd_t} \\
    \Delta\Sigma_{\times}(R, \psi) &= \frac{e}{2}\, Q_{\times}(R) \sin 2\psi ,
    \label{eq:esd_x}
\end{align}

where $\psi$ is the azimuthal angle measured from the halo major axis, $\Delta\Sigma_{0}(R)$ is the excess surface mass density of the spherical NFW profile \citep{Wright2000}, and $Q_{+}$ and $Q_{\times}$ are the quadrupole shape functions of \citet{Adhikari2015}. For the convergence we only keep the monopole term, $\kappa = \Sigma_{0}(R)/\Sigma_{\rm crit}$, the quadrupole contribution to $\kappa$ only affects the observed shapes of galaxies through the reduced-shear correction and is therefore a much smaller correction to the signal. Miscentering is modelled by displacing the halo centre by an offset $(\Delta x, \Delta y)$ with respect to the assumed cluster position, and the position angle $\phi_{0}$ sets the orientation of the major axis.

\subsection{Two-halo term}   

The two-halo term is the correlated large-scale structure (cLSS) surrounding galaxy clusters. Clusters live in very specific environments, the nodes of the cosmic web. In general more massive clusters live in denser environments \citep{Tinker2010}. Therefore the cLSS for more massive clusters contains a larger weak lensing signal. We neglect the covariance contribution from the clustering of haloes \citep[see][]{Wu2019}, which is only significant on scales larger than those considered here and, being independent of the number of clusters, is even less significant for individual-cluster lensing.

The azimuthal average of the cLSS weak-lensing signal for a cluster at a given mass can be calculated following \citet{oguri2011combining} and \citet{oguri2011detailed}. However we require the full two-dimensional field and crucially we need to understand the variance of this field, which is highly non-Gaussian.

We extract the two-halo contribution directly from numerical N-body simulations. We use the fiducial Quijote dark-matter-only boxes \citep{VillaescusaNavarro2020}, each with \(512^3\) particles in a \(1000\,h^{-1}\,\mathrm{Mpc}\) volume, and take the snapshot matching the lens redshift. We use the provided Rockstar catalogue to match the simulated cluster mass with the cluster mass we are simulating; a halo is selected at random from those within $0.2$ dex of the target mass \footnote{The interval of $0.2$ dex is chosen so that the mass constraints are only weakly dependent upon the cLSS}. The one-halo contribution is removed by subtracting the FoF member particles of the corresponding halo. The surface mass density is then estimated by projecting the remaining particles along the line of sight over a depth of approximately $300$ comoving Mpc, on a transverse patch of $90$ comoving Mpc centred on the halo. The transverse comoving extent is converted to an angular size at the lens redshift, and the surface mass density is converted to the convergence by dividing by the critical surface mass density (Eq. \ref{eq:convergence}).

The shear is then calculated from the convergence using the Kaiser-Squires inversion \citep{Kaiser1993}. 

\subsection{Uncorrelated large-scale structure}
\label{subsec:ulss}

In addition to the correlated structures surrounding the cluster, all structures along the line-of-sight will contribute to the lensing signal. This signal is essentially the cosmic shear signal, often used to constrain cosmology \citep{Asgari2021, Amon2022, Dalal2023, Li2023, Goh2026, Guerrini2026}. We model the uncorrelated large-scale structure (uLSS) as a Gaussian random field characterized by the convergence power spectrum $C_\ell^{\kappa\kappa}$. We calculate the power spectrum from the matter power spectrum $P_{\delta}(k, z)$ using the Limber approximation,

\begin{equation}
C_\ell^{\kappa\kappa} = \int_0^{\chi_{\rm h}} d\chi \, \frac{W^2(\chi)}{f_K^2(\chi)} P_{\delta}\left(\frac{\ell}{f_K(\chi)}, z(\chi)\right),
\end{equation}

where $\chi$ is comoving distance, $\chi_{\rm h}$ is the distance to the horizon, $n(z_{\rm s})$ is the source redshift distribution, $f_K(\chi)$ is the comoving angular diameter distance, and $W(\chi)$ is the lensing efficiency,

\begin{equation}
W(\chi) = \frac{3}{2} \Omega_{\rm m} H_0^2 \frac{f_K(\chi)}{a(\chi)} \int_\chi^{\chi_{\rm h}} d\chi' \, n(\chi') \frac{f_K(\chi' - \chi)}{f_K(\chi')},
\end{equation}
with $a(\chi)$ the scale factor and $n(\chi)$ the normalized source distribution.

We computed $C_\ell^{\kappa\kappa}$ using \texttt{CCL} \citep{Chisari2019} with a Planck 2018 cosmology \citep{Planck2020}. For the source distribution, we adopt a Smail-type profile,

\begin{equation}
n(z) \propto z^\alpha \exp\left[-\left(\frac{z}{z_0}\right)^\beta\right],
\end{equation}

with parameters $\alpha = 2.18$, $\beta = 1.07$, and $z_0 = 0.46$, which were chosen to match the $n(z)$ in the Euclid Flagship simulation \citep[][Appendix \ref{app:fs2}]{EuclidFlagship2025}. Additionally we set a minimum redshift $z_{\rm s} = 0.6$, below which we set $n(z) = 0$, to mimic the effect of photometric redshift cuts on the source sample.

To generate realisations of uLSS, we adopt a multi-scale approach to capture both small and large angular scales. Modes with $\ell \leq 500$ are generated on a coarse grid covering a $10$ degree field, then cropped and resampled onto the target grid, while modes with $\ell > 500$ are generated directly at the native resolution of the simulated field. The split at $\ell_{\rm cut} = 500$ is chosen to allow sufficient resolution for small scales, without making the generation of the random fields too slow. \citet{Murray2025a} show that the uLSS noise has contribution from very large scales.

\section{Simulations}
\label{sec:simulations}

The simulated shear fields are defined on a regular grid covering a $4$ degrees$^2$ field of view, with a pixel size of $15$ arcsec ($480\times480$ pixels).

We first calculate the two shear components $\gamma_1$, $\gamma_2$ and the convergence,

\begin{equation}
    \gamma_{i, \rm{total}} = \gamma_{i, \rm{1h}}(M, c , e , \phi, \Delta x, \Delta y) + \gamma_{i, \rm{cLSS}}(M) + \gamma_{i, \rm{uLSS}}
\end{equation}

\begin{equation}
    \kappa_{\rm{total}} = \kappa_{\rm{1h}}(M, c , e , \phi, \Delta x, \Delta y) + \kappa_{\rm{cLSS}}(M) + \kappa_{\rm{uLSS}}
\end{equation}

From which we calculate our observable, the reduced shear (Eq.~\ref{eq:reduced_shear}), $g_{i} = \gamma_{i, \rm{total}} / (1-\kappa_{\rm{total}})$.

Within our forward model we can at this point then include the effects of dilution by foreground and cluster member galaxies and the intrinsic alignment of cluster members around the cluster. This has been implemented within the code, however we do not study them in this paper. The amplitude of both effects depends strongly on the quality of the photometric redshifts and on the source selection of a particular survey. Therefore any value we chose would describe a particular survey rather than a property of galaxy clusters. We set both to zero throughout this work.

\begin{figure}[htb]
    \centering
    \includegraphics[width=\columnwidth]{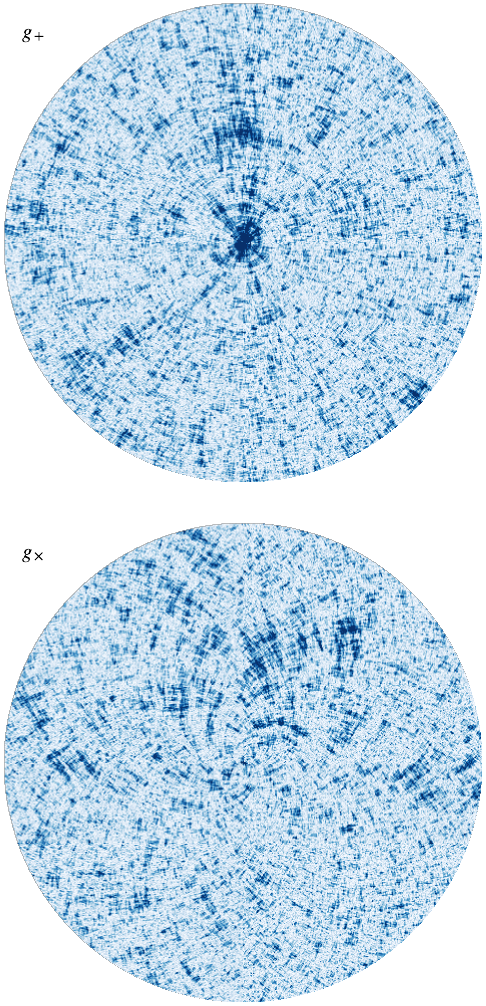}
    \caption{Simulated reduced-shear field around a cluster with $M_{200m} = 8\times10^{14}\,{\rm M}_\odot$, $c = 4$, $e = 0.4$, miscentering $0.2$ arcmin and $z_{\rm lens} = 0.5$, without shape noise. \emph{Top:} tangential component $g_+$. \emph{Bottom:} cross component $g_\times$.}
    \label{fig:shear_fields}
\end{figure}

Finally we add the galaxy shape noise assuming an intrinsic ellipticity dispersion per component of $\sigma_\epsilon = 0.26$. The noise contribution within each pixel is drawn from a Gaussian distribution with variance $\sigma_\epsilon^2 / N_{\rm gal, pix}$, where $N_{\rm gal, pix}$ is the number of galaxies in the pixel. An example of a single realisation is shown in Figure \ref{fig:shear_fields}, without shape noise so that the signal is visible on the unsmoothed field.

\subsection{Simulation-based inference}
\label{subsec:sbi}

The simulations described in the previous section produce cluster fields with non-Gaussian and correlated noise across the field. Writing down an explicit likelihood for the fields would require the joint estimation of the cluster parameters, the non-Gaussian correlated large-scale structure, and the uncorrelated line-of-sight structure. This is the approach taken for field-level inference \citep[e.g. BORG,][]{Jasche2013}. Alternatively we could attempt to model the full pixel-to-pixel covariance, but in doing so we would not capture the non-Gaussianity of the correlated large-scale structure. Using simulation-based inference we can instead generate many realisations of the data and learn the posterior directly from these simulations \citep{Cranmer2020}. This allows us to perform accurate inference on these complicated fields. 

\begin{figure}[htb]
    \centering
    \includegraphics[width=\columnwidth]{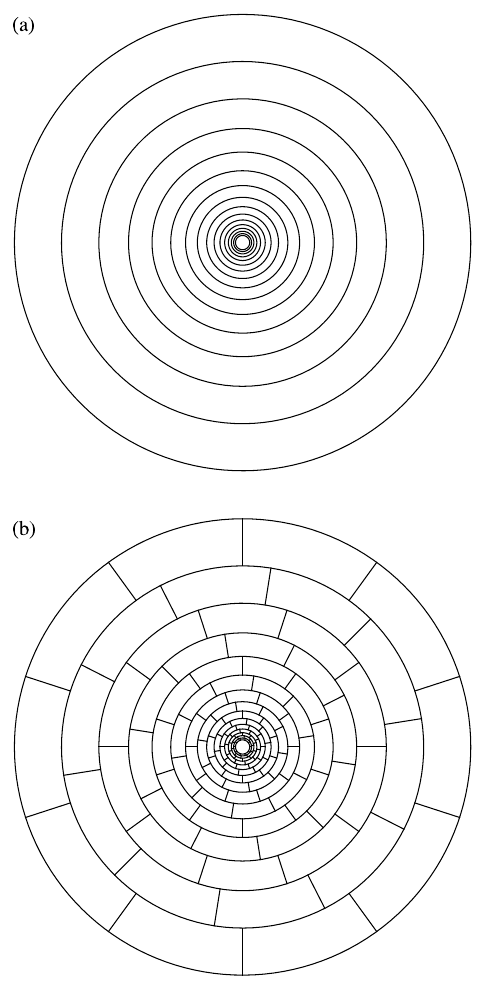}
    \caption{Geometry of the two summary statistics. \emph{(a)} 1D summary: 15 logarithmically spaced annuli between $1$ and $32$ arcmin. \emph{(b)} 2D summary: the same annuli, each divided into 10 angular sectors.}
    \label{fig:summary_statistics}
\end{figure}

The simulations produce high-resolution maps of the shear field, however we prefer not to use this information directly, as the small scales of the shear field are challenging to model and we do not have confidence that this can currently be done correctly. Therefore we compress the shear field into summary statistics. The geometry of the fiducial summary statistics is shown in Figure \ref{fig:summary_statistics}. For each simulation, we generate a one-dimensional summary statistic by azimuthally averaging the shear in 15 logarithmically spaced radial bins between $1$ and $32$ arcmin. The two-dimensional summary statistic uses the same radial bins, but each annulus is further separated into 10 angular sectors. Information at scales smaller than one arcminute is not used in this work, as we expect the signal at such scales to be poorly modelled by the NFW profile. Both the tangential and the cross-shear components are used within the summary statistics. The $g_+$ and $g_\times$ measurements are concatenated together, giving 30 elements for the 1D summary and $15\times10\times2 = 300$ for the 2D summary. In practice we pass our summary statistics through a learned embedding network, as explained below. The impact of using only the tangential component is
quantified in Section~\ref{subsec:cross_shear}, and the sensitivity of the inference to the number of radial bins in Appendix~\ref{app:resolution}.

For this work we used the \texttt{sbi} Python package \citep{Tejero2020}, which implements simulation-based inference using neural density estimation with conditional normalizing flows \citep{Papamakarios2019, Hermans2020}. We use neural posterior estimation \citep{Greenberg2019} with a masked autoregressive flow \citep{Papamakarios2017} as the conditional density estimator. 

For the 2D approach, the input to the density estimator is the concatenation of the 1D and 2D summaries, 330 elements. The 1D statistic contributes no new information to the 2D statistic, however we found that including it slightly improved the precision on the cluster mass. This is understandable as the mass information is mostly contained within the 1D profile and there is little parameter degeneracy between the mass and the other parameters. Before the flow, the summary statistics are passed through a learned embedding network which maps them to 64 features. This is essential for the 2D summary, as the flow struggles to learn the posterior from high-dimensional inputs. The compression improves the convergence of the flow and the accuracy of the posterior \citep{Alsing2018, Charnock2018}.

For each of the different setups, we generate a set of $10^6$ simulations\footnote{$10^6$ simulations are used to train the posteriors for each method, however throughout the results section different numbers of simulations are used to test the noise contributions and evaluate the constraining power of each method.}. The cluster mass prior is uniform in $\log_{10}(M_{200m}/M_\odot)$ over the range $[13.6,\,15.4]$. The concentration prior is uniform over $c\in[0.1,\,15.0]$. The cluster ellipticity prior is uniform over $e\in[0,\,0.8]$, and the position angle prior is uniform over $\phi\in[0^\circ,\,180^\circ)$. Miscentering offsets are drawn uniformly over the disc of radius $1$ arcmin (uniform in area, so the radial offset satisfies $\Delta R = \sqrt{\Delta x^{2} + \Delta y^{2}} \leq 1$ arcmin with prior mean $\langle \Delta R \rangle = 2/3$ arcmin). Since the expensive forward model is independent of the shape-noise level, we compute the noiseless summaries once and add shape noise at the summary level, allowing the same simulation set to be reused across source densities.

\section{Results}
\label{sec:results}

With our simulation framework in place we first explore the different noise contributions to one-dimensional shear profiles, and then compare the performance of one-dimensional and two-dimensional summary statistics for cluster parameter inference.

\subsection{Noise contributions}
\label{subsec:noise_overviews}

We assess the different noise contributions to the cluster shear profiles by running simulations with different effects turned on and off. 

For each case we generate 5000 simulations, with the fiducial parameters $M_{200m} = 10^{15}$ M$_\odot$, $c_{200m} = 4.0$, ellipticity $e = 0.0$, $z_{\rm lens} = 0.5$. Both the uLSS and cLSS contributions are generated as described in Section \ref{sec:simulations}. The uLSS contribution depends strongly on the source redshift distribution, as higher redshift sources accumulate considerably more uLSS noise than low-redshift sources, simply because they have had more occasion to be lensed by foreground massive objects. The miscentering contribution is created by scattering the cluster centre by a random offset drawn from a Gaussian with $\sigma(\Delta x, \Delta y) = 0.4$ arcmin, which should be typical miscentering for optical cluster detection \citep[e.g.][]{Simet2017, Melchior2017, McClintock2019}. The concentration is scattered with $\sigma(\ln c) = 0.33$, the intrinsic scatter at fixed mass measured for cluster-scale haloes by \citet{Bhattacharya2013} \citep[see also][]{Diemer2019}. We draw the cluster ellipticity from a Gaussian with mean $0.4$ and $\sigma(e) = 0.15$, clipped to $[0, 0.8]$ \citep[consistent with the intrinsic shape distribution of cluster-mass haloes,][]{Despali2017}, with the position angle sampled uniformly on $[0^\circ, 180^\circ)$.

From the $N$ realisations of each case we compute the covariance matrix of the binned tangential shear,
\begin{equation}
C_{ij} = \frac{1}{N-1} \sum_{k=1}^{N} \left[ g_{t,k}(R_i) - \bar{g}_t(R_i) \right] \left[ g_{t,k}(R_j) - \bar{g}_t(R_j) \right],
\label{eq:covariance}
\end{equation}
where $g_{t,k}(R_i)$ is the azimuthally averaged tangential shear of realisation $k$ in radial bin $i$ and $\bar{g}_t(R_i)$ is its mean over realisations.

\begin{figure}[htb]
    \centering
    \includegraphics[width=\columnwidth]{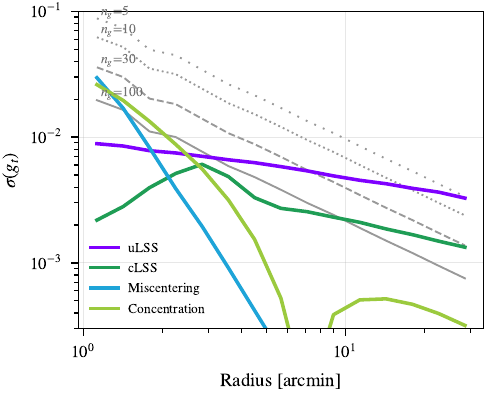}
    \caption{Contributions to the standard deviation $\sigma(g_t)$ of the azimuthally averaged reduced tangential shear, from 5000 realisations per setup. Grey dotted lines: shape noise at source densities $n_g = 5$, $10$, $30$ and $100\,{\rm arcmin}^{-2}$.}
    \label{fig:gt_variance}
\end{figure}

Figure~\ref{fig:gt_variance} shows each noise contribution to the standard deviation of the measured tangential shear profile, with the shape-noise contribution shown for a range of source densities. On small scales, the noise is dominated by shape-noise except for very high source densities ($n_g=100$). On small scales miscentering and concentration scatter are more important than uLSS and cLSS. This is in agreement with the results of \cite{Gruen2015}. At larger scales (around 7 arcminutes for a density of 30 per square arcminute \footnote{This is the expected source density for Stage IV surveys; Stage III surveys reach $6$--$20\,{\rm arcmin}^{-2}$.}), the uLSS contribution becomes dominant compared to shape-noise. The cLSS contribution is substantially smaller at all scales. However it is strongly correlated between radial bins, and as we show in Section~\ref{subsec:clss_results} it still increases the mass uncertainty of the most massive clusters by $16\%$. The result is also in agreement with \cite{Wu2019}, where the 1-halo term (cluster profile variations) become important relative to the uLSS at small-scales. The halo ellipticity is not shown in Figure~\ref{fig:gt_variance}, as it does not contribute to the 1D profile. The quadrupole of Eq.~\ref{eq:esd_t} averages to zero over a full annulus.

Figure~\ref{fig:gt_variance} also indicates where we expect the 2D summary to help. Miscentering and concentration scatter dominate the noise in the inner few arcminutes, and the signal from these effects is anisotropic, therefore these are the parameters for which we expect the 2D summary to gain the most. 

\subsection{Mass, concentration, miscentering and ellipticity constraints}
\label{subsec:mass_results}

Now we compare the constraints on the cluster parameters from the 1D and 2D methods at different levels of shape noise.

\begin{figure*}[htb]
    \centering
    \includegraphics[width=18cm]{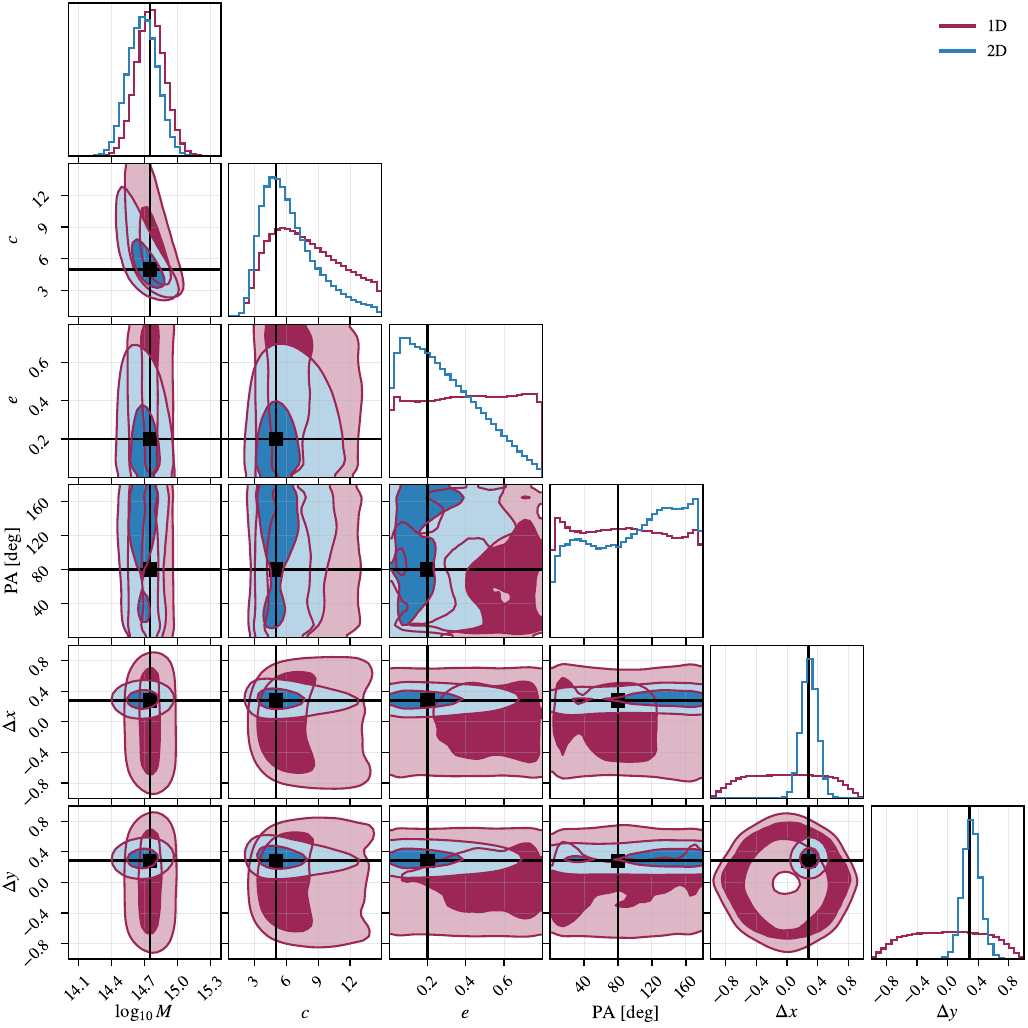}
    \caption{Posteriors from the 1D (purple) and 2D (blue) summaries for a single noise realisation of one cluster at $n_{\rm gal} = 100\,{\rm arcmin}^{-2}$, with $1\sigma$ and $2\sigma$ contours and $3\times10^{6}$ samples each. Black lines: true parameter values.}
    \label{fig:posterior_comp}
\end{figure*}

Firstly we apply both methods to a single realisation of a cluster shear field, at $n_{\rm gal} = 100\,{\rm arcmin}^{-2}$ so that the differences between the two summaries are at their clearest. The posterior from this is shown in Figure \ref{fig:posterior_comp} and the 1-$\sigma$ constraints on each parameter, for both summaries, are summarised in Table~\ref{tab:single_cluster_constraints}. For this particular cluster realisation the mass constraints are similar for both methods, but the 2D method leads to significant improvement for the concentration, ellipticity and miscentering. 

The largest improvement is for the cluster centering. This is unsurprising, as miscentering generates a highly anisotropic shear signal in the coordinates of the assumed cluster centre. Within the 1D profile this will simply appear as a decrement in the tangential shear, which leads to a small degeneracy with the cluster concentration as seen in the posterior. It might be expected that the improvement in concentration and miscentering would in turn improve the constraints on the cluster mass. However we see that there is little degeneracy between these parameters. As shown in \citet{Murray2022}, the cluster mass is primarily constrained by the amplitude of the tangential shear at intermediate radii. These intermediate radii are largely insensitive to both cluster concentration and miscentering, therefore the improvements in miscentering and concentration do not translate into improvements on the cluster mass estimation.

\begin{table}[htb]
    \centering
    \caption{Marginal $1\sigma$ constraint on each cluster parameter from the 1D and 2D summaries, for the single realisation of Fig.~\ref{fig:posterior_comp}.}
    \label{tab:single_cluster_constraints}
    \begin{tabular}{lcc}
\hline\hline
Parameter & 1D & 2D \\
\hline
$\sigma(\log_{10} M_{200m})$ & 0.129 & 0.132 \\
$\sigma(c_{200m})$ & 3.24 & 2.66 \\
$\sigma(e)$ & 0.229 & 0.188 \\
$\sigma(\phi)$ [deg] & 51.4 & 49.1 \\
$\sigma(\Delta x)$ [arcmin] & 0.478 & 0.104 \\
$\sigma(\Delta y)$ [arcmin] & 0.471 & 0.116 \\
$\sigma(\Delta R)$ [arcmin] & 0.204 & 0.111 \\
\hline
\end{tabular}

\end{table}

The posteriors of individual cluster realisations have a variance associated to the particular realisation of the field. In order to overcome this, we simulate many cluster realisations and each time obtain the posterior. We then compare the constraining power of the different summary statistics using the mean of the $1\sigma$ posterior widths over the many realisations,

\begin{equation}
\langle \sigma(\theta) \rangle = \frac{1}{K} \sum_{k=1}^{K} \sigma_{{\rm post},k}(\theta) ,
\label{eq:mean_sigma}
\end{equation}

where $\sigma_{{\rm post},k}(\theta)$ is the standard deviation of the marginal posterior for parameter $\theta$ in noise realisation $k$, and $K = 800$ realisations are used per grid point in Figs.~\ref{fig:param_grid} and \ref{fig:cross_shear_impact}, $K = 200$ in Fig.~\ref{fig:clss_impact} and in Appendix~\ref{app:resolution}. At each point of the grid one parameter is fixed and the other parameters are drawn from the prior for each realisation. Therefore $\langle \sigma(\theta) \rangle$ is the marginalised uncertainty that would be obtained for a single cluster. We use the same metric in Figures~\ref{fig:param_grid}, \ref{fig:cross_shear_impact} and \ref{fig:clss_impact} so that they can be compared directly. We confirmed the accuracy of our posteriors using coverage tests, which are presented in Appendix~\ref{app:coverage}.

The results are shown in Fig.~\ref{fig:param_grid}. The 2D summary gives smaller errors than the 1D summary for every parameter, at every point of the grid and every source density, and the advantage grows as the shape noise decreases.

The 2D mass estimation has consistently smaller errors than the 1D mass estimation, with an improvement of between $3$ and $18\%$. The largest improvements are at $n_{\rm gal} = 100\,{\rm arcmin}^{-2}$, and the improvement is $8\%$ on average over the mass grid at $30\,{\rm arcmin}^{-2}$. The 2D summary can weight the correlated noise across the field more optimally than the azimuthal average (see \citealt{Murray2025a}), this is investigated in Subsec~\ref{subsec:clss_results}.

The miscentering produces an anisotropic shear signal, therefore it is almost invisible to the azimuthal average, which only sees a decrement of the tangential shear at small radii. This can be seen directly in the posterior widths. The miscentering panel shows the radial offset $\Delta R = \sqrt{\Delta x^{2} + \Delta y^{2}}$, for which the prior (Section~\ref{subsec:sbi}) has a standard deviation of $0.236$ arcmin. The 1D width is between $0.18$ and $0.22$ arcmin for every true offset and every source density, which is close to the prior. The 2D error is below the 1D error at every point and decreases as the true offset increases, from $0.20$ to $0.18$ arcmin at $n_{\rm gal} = 10\,{\rm arcmin}^{-2}$ and from $0.15$ to $0.11$ arcmin at $100\,{\rm arcmin}^{-2}$. Therefore the improvement relative to 1D increases with both the source density and the offset. The improvement is $21\%$ on average at $30\,{\rm arcmin}^{-2}$, rising to $32\%$ at $100\,{\rm arcmin}^{-2}$, and it ranges between $8$ and $38\%$ across the whole grid.

The concentration depends on the shear profile at small radii, as can be seen in Figure~\ref{fig:gt_variance}, where varying the concentration primarily leads to variance at small radii. Because of this the concentration estimation depends more strongly on the shape noise than the mass estimation does, as the inner annuli contain the fewest source galaxies. Between $n_{\rm gal} = 10$ and $100\,{\rm arcmin}^{-2}$ the 1D error decreases by only $11\%$, whereas the 2D error decreases by $23\%$. Therefore the advantage of the 2D summary grows from $7\%$ to $19\%$ over this range, and is $11\%$ at $30\,{\rm arcmin}^{-2}$.

The ellipticity is poorly constrained for an individual cluster, as also found by \citet{Payerne2024} from a multipole analysis of cluster shear profiles in The Three Hundred simulations. The 1D error is between $0.226$ and $0.228$ at every grid point and every density, which is within $2\%$ of the standard deviation of the uniform prior, $0.231$. The 2D summary is only $\sim 2\%$ tighter than 1D at $n_{\rm gal} = 10\,{\rm arcmin}^{-2}$, however it does extract ellipticity information as the shape noise decreases, with an improvement of $6$--$8\%$ at $30\,{\rm arcmin}^{-2}$ and $10$--$13\%$ at $100\,{\rm arcmin}^{-2}$. The position angle is also unconstrained for a single cluster (Table~\ref{tab:single_cluster_constraints}). Therefore the ellipticity constraints for individual clusters remain prior dominated at Stage IV depths. The same forward model can however be applied to stacked cluster samples, for which the prior will no longer dominate.

\begin{figure*}[htb]
    \centering
    \includegraphics[width=\textwidth]{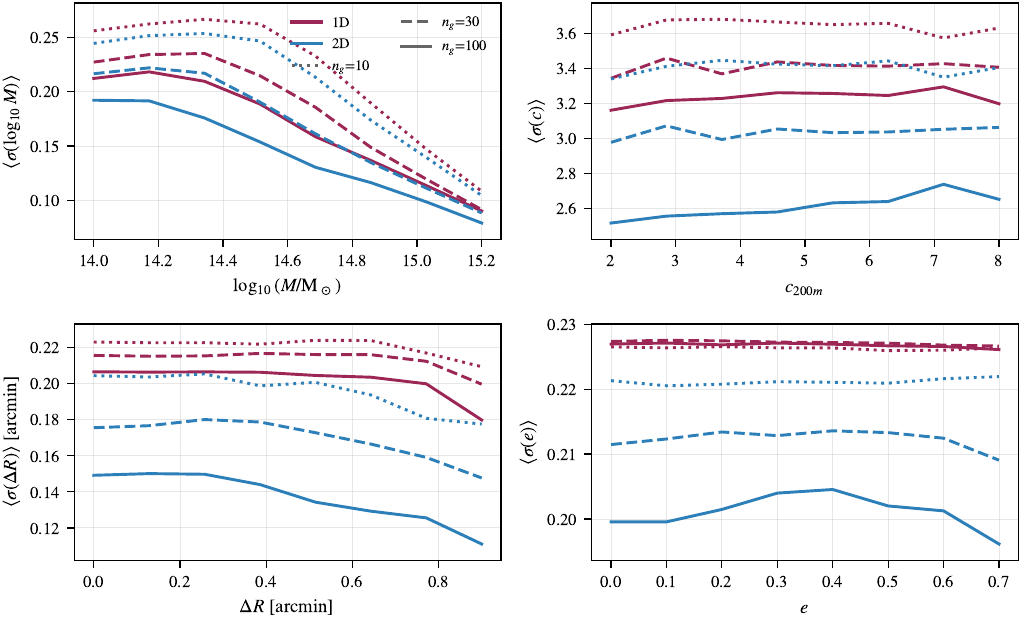}
    \caption{Mean marginal posterior width $\langle \sigma(\theta) \rangle$ (Eq.~\ref{eq:mean_sigma}) for $\log_{10}M$, $c$, the radial miscentering $\Delta R$ and the ellipticity $e$, over $K = 800$ noise realisations per grid point. Red: 1D summary; blue: 2D summary. Dotted, dashed and solid lines: $n_{\rm gal} = 10$, $30$ and $100\,{\rm arcmin}^{-2}$.}
    \label{fig:param_grid}
\end{figure*}

\subsection{Impact of the cross-shear component}
\label{subsec:cross_shear}

The fiducial summary statistics contain both tangential and cross components of the reduced shear. For a circularly symmetric lens, the cross component carries no signal and is therefore discarded in most cluster lensing analyses (or used only as a systematic null test). However, both the halo ellipticity and the miscentering create a coherent cross-shear signal. We determined the usefulness of the cross-shear signal by creating two more summary statistics for the source density, $n_{\rm gal} = 30\,{\rm arcmin}^{-2}$, containing only the $g_+$ measurements. All four posteriors are trained on the same set of simulations and applied to the same set of mock observations, so that the only difference between them is the cross-shear component.

\begin{figure*}[htb]
    \centering
    \includegraphics[width=\textwidth]{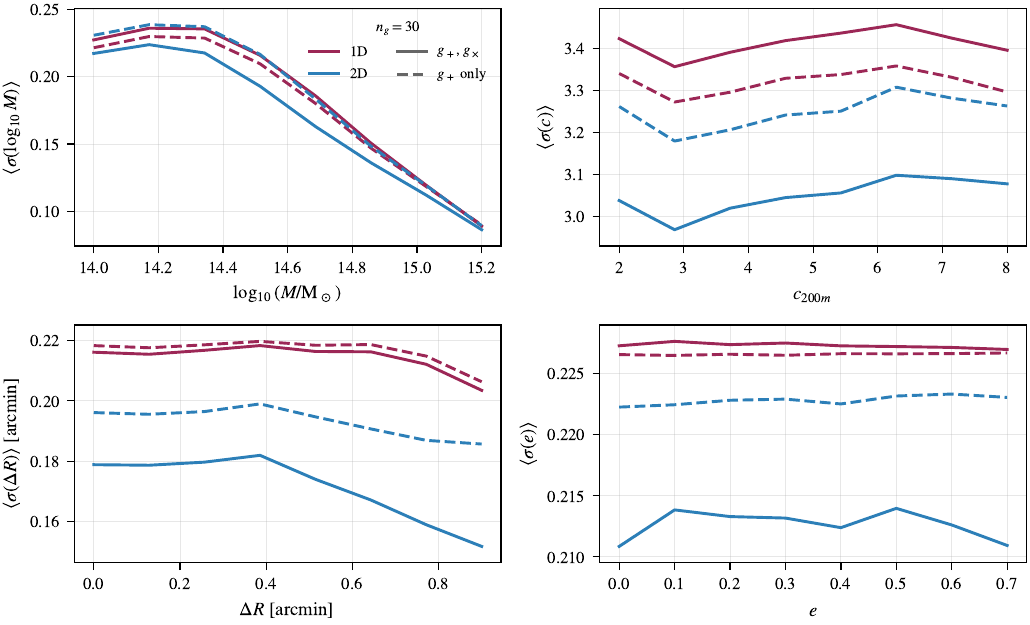}
    \caption{Mean marginal posterior width $\langle \sigma(\theta) \rangle$ (Eq.~\ref{eq:mean_sigma}) at $n_{\rm gal} = 30\,{\rm arcmin}^{-2}$ with and without the cross-shear component, over $K = 800$ noise realisations per grid point with $4000$ posterior samples each. Red: 1D summary; blue: 2D summary. Solid: $g_+$ and $g_\times$; dashed: $g_+$ only.}
    \label{fig:cross_shear_impact}
\end{figure*}

These results are shown in Figure \ref{fig:cross_shear_impact}. The cross-shear information, has no impact for the 1D summary statistic. However the cross-shear improves respectively the miscentering and concentration estimation by up to $18\%$ and $7\%$ (the largest reduction in $\langle\sigma(\theta)\rangle$ across the grid, reached at $\Delta R = 0.9\,{\rm arcmin}$ and $c_{200m} = 2$). The improvement for the ellipticity estimation is smaller, at most $5\%$; this reflects that the ellipticity is only poorly constrained for individual clusters. 

\subsection{Impact of the uncorrelated and correlated large-scale structures}
\label{subsec:clss_results}

\begin{figure*}[htb]
    \centering
    \includegraphics[width=\textwidth]{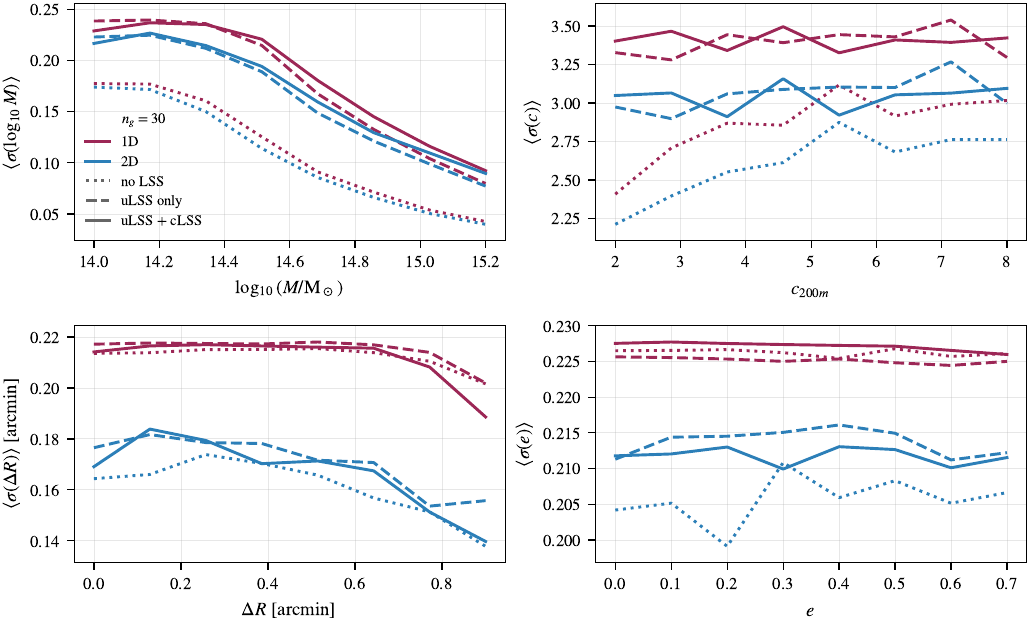}
    \caption{Mean marginal posterior width $\langle \sigma(\theta) \rangle$ (Eq.~\ref{eq:mean_sigma}) at $n_{\rm gal} = 30\,{\rm arcmin}^{-2}$ for three forward models, over $K = 200$ realisations per grid point. Red: 1D summary; blue: 2D summary. Dotted: no line-of-sight structure; dashed: uLSS only; solid: uLSS and cLSS (the fiducial model).}
    \label{fig:clss_impact}
\end{figure*}

We now consider the impact of uncorrelated and correlated large-scale structure on the parameter constraints. We compared three types of model; the fiducial model used throughout the paper, a model without the cLSS component, and a model without cLSS or uLSS. For each model the posterior is trained and tested on simulations from that same model. Therefore we can look at directly the change in the obtained posterior widths, rather than worrying about model specification. We remove the contributions one at a time in order to see which of the contributions is responsible for the loss in constraining power. These results are shown in Fig~\ref{fig:clss_impact}.

The fiducial model (uLSS+cLSS) can be considered to be the true error bars, anything that gives tighter constraints illustrates that something is missing in the modelling. In Fig~\ref{fig:clss_impact} we see that the estimated uncertainty of the mass estimation for no LSS is far too small. At large masses this approaches a factor of two, the no-LSS width being $2.2$ times smaller than the fiducial one at $\log_{10}(M_{200m}/{\rm M}_\odot) = 15.2$, falling to a factor $1.3$ at $10^{14}\,{\rm M}_\odot$. Between uLSS+cLSS and uLSS only the difference is much smaller, reaching $16\%$ for both summaries for the largest mass clusters, and the difference goes to zero at masses of $10^{14}$. Therefore the uLSS is responsible for most of the loss of constraining power, as expected from Figure~\ref{fig:gt_variance}. Similar but less significant trends are seen for the concentration, miscentering and ellipticity.

The same three models also tell us about the difference between the two summaries. The difference between the red and blue curves is the advantage of the 2D approach within a fixed forward model. This allows us to separate the two explanations for the mass improvement, discussed in Section~\ref{subsec:mass_results}. When no LSS is used in the forward model, the improvement between mass constraints between the 1D and the 2D approach is minimal, whilst the constraints on concentration are still significant. This shows that the mass constraint improvement for the 2D approach is from a better treatment of the spatially correlated LSS noise across the cluster field (discussed in detail in \cite{Murray2025a}).

\section{Discussion and conclusions}
\label{sec:discussion}

In this paper we introduced a method to estimate cluster mass, concentration, miscentering and ellipticity using the two-dimensional weak lensing shear field. By forward modelling the field we are able to create a complex and high fidelity model which includes: uncorrelated large-scale structure, the non-Gaussian correlated large-scale structure, the reduced shear, miscentering and cluster ellipticity. We created two different summary statistics, 1D and 2D, and trained a neural posterior estimator on each. This allowed us to assess the improvements from the two-dimensional information.

The 2D summary improves the constraints on every parameter, and the improvement grows as the shape noise decreases. At a Stage IV source density of $30\,{\rm arcmin}^{-2}$ the improvements are $8\%$ for the mass, $11\%$ for the concentration and $21\%$ for the miscentering, rising to $18$, $19$ and $32\%$ at $100\,{\rm arcmin}^{-2}$ (Fig.~\ref{fig:param_grid}). The cross-shear signal contributes to these improvements, as it contains information on the anisotropic parameters. Including it improves the miscentering and concentration constraints by up to $18$ and $7\%$ respectively (Sect.~\ref{subsec:cross_shear}). The ellipticity of individual clusters however remains prior dominated. The 1D summary is prior dominated for the ellipticity, and the 2D summary improves on the prior by only $6$--$8\%$ at $30\,{\rm arcmin}^{-2}$ and $10$--$13\%$ at $100\,{\rm arcmin}^{-2}$.

We also quantified the impact of the line-of-sight structure on the constraints. Neglecting it underestimates the mass uncertainty by a factor which grows with the mass and reaches $2.2$ at $\log_{10}(M_{200m}/{\rm M}_\odot) = 15.2$. This is primarily from the uLSS. The cLSS adds a further $16\%$ at the highest masses, and has no effect at $10^{14}\,{\rm M}_\odot$ (Fig.~\ref{fig:clss_impact}). 

Coverage tests show that the posteriors are correctly calibrated (Appendix~\ref{app:coverage}), and the forward model reproduces the mean shear profiles of the Euclid Flagship simulation on the scales used in the inference, and the line-of-sight noise on all scales (Appendix~\ref{app:fs2}).

A limitation of the current model is the absence of cluster substructures. This could be one reason for the discrepancy between the small-scale variance of our shear profiles and those from the Euclid Flagship simulation (discussed in App.~\ref{app:fs2}).

A second limitation is that we have not studied the dilution of the source sample by foreground and cluster member galaxies, or the intrinsic alignment of the cluster members with the halo. Both of these effects are included in the forward model and can be found in the provided \verb|github| repository, however they are strongly dependent on the survey. The dilution can be close to zero with high quality photometric redshifts, and large when foreground and background galaxies are confused. The natural way to treat these effects is within the forward model, rather than by correcting the measured profiles afterwards.

Once the method has been trained, posterior inference is extremely fast per cluster. For example $4000$ posterior samples for a cluster takes $\sim 0.1$ seconds on a single laptop CPU core. This makes the approach practical for large cluster samples from Euclid \citep{Laureijs2011} and LSST \citep{LSST2009}. However adapting the method to the variable source density across a survey and to survey masks will require some additional work, as the forward model must then be conditioned on the noise properties of each cluster field.

\begin{acknowledgements}
The initial version of the code and analysis was written by CM. The code was subsequently rewritten with the assistance of large language models under CM's direction. CM wrote the manuscript, with comments from MK and CP; large language models were used for language editing. The analysis made use of ASTRA (Agentic Schema for Transparent Research Analysis) and \texttt{lightcone-cli}, both developed by Lightcone Research (\url{https://github.com/LightconeResearch}). The code will be made available on publication of the article \url{https://github.com/calumhrmurray/two_dimensional_lensing_lc}.

This analysis was conducted on the CANDIDE cluster at the Institut d’Astrophysique de Paris. The cluster is funded through grants from the PNCG, CNES, DIM-ACAV, the Euclid Consortium, and the Danish National Research Foundation Cosmic Dawn Center (DNRF140). It is maintained by Stephane Rouberol. 

This work made use of the Quijote simulations \citep[][\url{https://quijote-simulations.readthedocs.io}]{VillaescusaNavarro2020}, which we accessed through the Quijote Binder hosted at the Flatiron Institute. We also made use of the Euclid Flagship simulation \citep{EuclidFlagship2025}, which we accessed through CosmoHub \citep{Carretero2017, Tallada2020}. CosmoHub has been developed by the Port d'Informaci\'o Cient\'ifica (PIC), maintained through a collaboration of the Institut de F\'isica d'Altes Energies (IFAE) and the Centro de Investigaciones Energ\'eticas, Medioambientales y Tecnol\'ogicas (CIEMAT) and the Institute of Space Sciences (CSIC \& IEEC), and was partially funded by the ``Plan Estatal de Investigaci\'on Cient\'ifica y T\'ecnica y de Innovaci\'on'' program of the Spanish government.
\end{acknowledgements}
 
\bibliographystyle{aa}
\bibliography{references}

\appendix

\section{Comparison with Euclid Flagship simulations}
\label{app:fs2}

In this Appendix we compare our forward model against the Euclid Flagship simulation \footnote{Downloaded from the CosmoHub website: \url{https://cosmohub.pic.es/home}} \citep[FS2,][]{EuclidFlagship2025}. We selected clusters in the redshift interval $0.45 \leq z \leq 0.55$ and then separated them into three mass bins, $\log_{10}(M_{200m}/{\rm M}_\odot) \in [14.0,\,14.3]$, $[14.3,\,14.7]$ and $[14.7,\,15.3]$, which contain $427$, $133$ and $5$ haloes respectively, on a subregion of FS2 of $\sim 180$ deg$^2$. We then measured the stacked shear profiles for each mass bin using \texttt{TreeCorr} \citep{Jarvis2004} \footnote{\url{https://rmjarvis.github.io/TreeCorr/_build/html/overview.html}}. For each mass bin we simulated $300$ cluster realisations with the same mass and redshift distribution. The source redshifts follow a Smail $n(z)$ fit to the FS2 source distribution (Sect.~\ref{subsec:ulss}).

\begin{figure}[htb]
    \centering
    \includegraphics[width=\columnwidth]{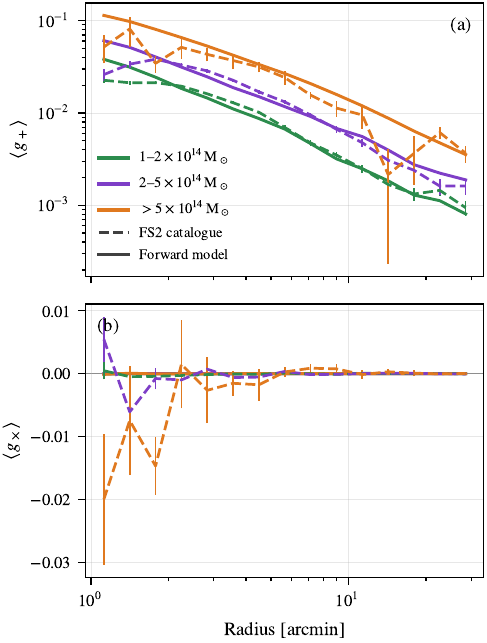}
    \caption{Mean tangential \emph{(a)} and cross \emph{(b)} reduced shear profiles in three mass bins, for the forward model (solid lines) and the stacked FS2 haloes (dashed lines with points).}
    \label{fig:fs2_profiles}
\end{figure}

In Figure~\ref{fig:fs2_profiles} we show the comparison between the forward model and the FS2 measurements. Beyond $\sim 3$ arcmin the agreement is good in all three mass bins, and as expected the cross-shear signal is consistent with zero for each of the measurements. Inside $\sim 3$ arcmin the FS2 profiles flatten while the model continues to rise, reaching a factor of $\sim 2$ in the innermost bin. This is an issue of resolution of the FS2 weak lensing field. The FS2 lensing fields are obtained by projecting the mass onto all-sky HEALPix maps at $N_{\rm side} = 8192$, which has a pixel scale of $0.43$ arcmin, whereas our two innermost annuli are only $0.26$ and $0.33$ arcmin wide. Therefore the inner profile is smoothed by the map pixelisation, and the two curves converge at the radius where the annulus width becomes larger than the FS2 pixel. We restrict the quantitative comparison below to radii beyond this scale.

\begin{figure}[htb]
    \centering
    \includegraphics[width=\columnwidth]{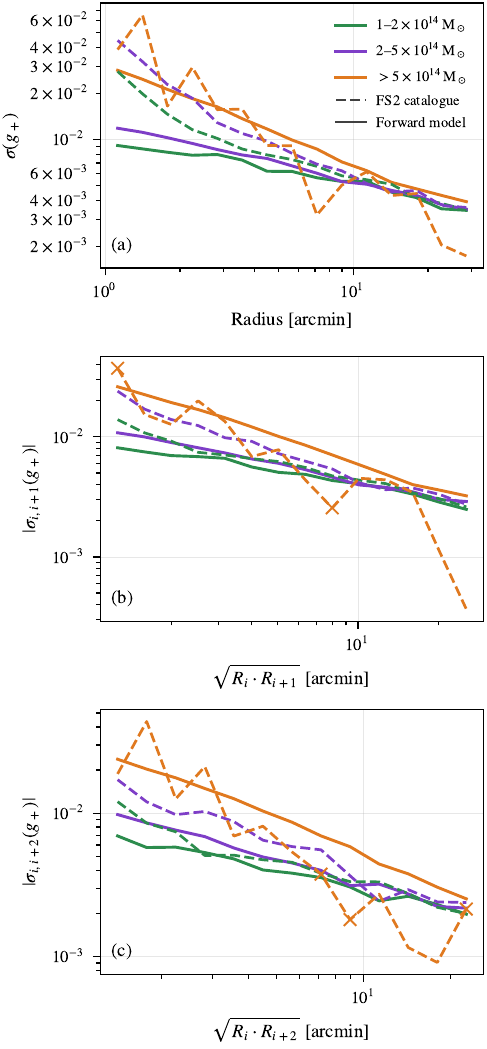}
    \caption{Standard deviation of the tangential shear profile \emph{(a)} and the magnitude of the first \emph{(b)} and second \emph{(c)} off-diagonal terms of its covariance, plotted against the geometric mean radius of the two bins, in three mass bins for the forward model (solid lines) and FS2 (dashed lines).}
    \label{fig:fs2_covariance}
\end{figure}

\begin{figure}[htb]
    \centering
    \includegraphics[width=\columnwidth]{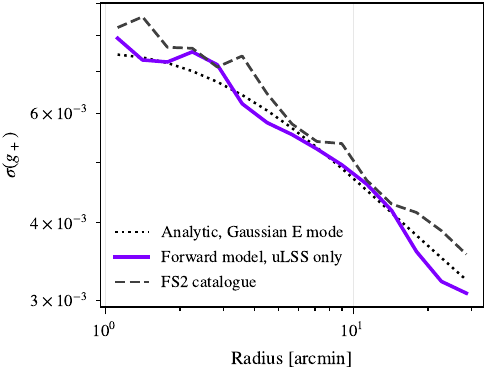}
    \caption{Standard deviation of the tangential shear profile from uLSS alone, in 15 logarithmically spaced annuli between $1$ and $32$ arcmin, from the forward model (purple), the analytic Gaussian prediction (black dotted) and 300 random pointings in FS2 (grey dashed).}
    \label{fig:ulss_random}
\end{figure}

We then compared the standard deviation of our profiles to those from FS2. This is a non-trivial test of our noise model. In Fig.~\ref{fig:fs2_covariance} we compare the noise in our model to those observed in the FS2 profiles. For the diagonal terms we see that FS2 has larger variance at small radii. We believe that this can be explained by substructures in clusters, which are not included within our model. For the radial annuli substructures will produce shot-noise. In support of the shot-noise hypothesis we see that the off-diagonal terms in FS2 are in better agreement with our model (shot-noise contributes only to the diagonal terms). This idea is further supported by the close agreement between our model and FS2 in the uLSS regime where cluster substructure is not important, see Fig. \ref{fig:ulss_random} and the discussion of this Figure in the following paragraph. 

In order to validate our uLSS noise we can compare the forward model to FS2 and the analytical prediction (see \citep{Hoekstra2003,1992Schneider,Murray2025a}). We measure uLSS in FS2 by taking 300 randoms points within FS2 and measuring the standard deviations of the shear profiles. The agreement between the three different measurements is excellent. 

Throughout this appendix we use FS2 as an independent, fully non-linear realisation of cluster lensing, rather than as the truth. FS2 has a finite mass and force resolution, its lensing fields are pixelised at the $0.43$ arcmin scale discussed above therefore we should not expect or look for perfect agreement with FS2. 

\section{Varying the number of radial bins}
\label{app:resolution}

\begin{figure*}[htb]
    \centering
    \includegraphics[width=\textwidth]{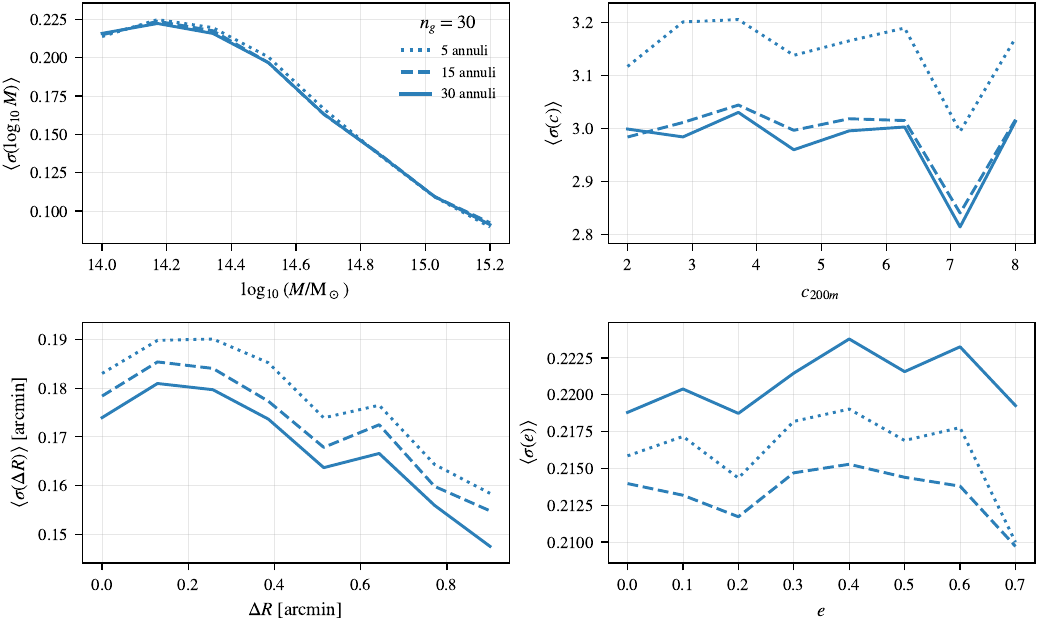}
    \caption{Mean marginal posterior width $\langle \sigma(\theta) \rangle$ (Eq.~\ref{eq:mean_sigma}) for the 2D summary at $n_{\rm gal} = 30\,{\rm arcmin}^{-2}$ with 5 (dotted), 15 (dashed) and 30 (solid) radial bins, over $K = 200$ realisations per grid point.}
    \label{fig:resolution_sigma}
\end{figure*}

To verify that our results are general and do not depend too strictly on the number of radial bins we repeat the results of Subsection \ref{subsec:mass_results} using 5,15 and 30 logarithmically spaced bins. This uses the same set as simulations as the fiducial analysis except we rebin the 2D summary statistics. Therefore the noise realisation is consistent across the different bin resolutions. The dependence on the number of radial bins is weak for the mass estimation, with $\langle\sigma(\log_{10}M)\rangle$ changing by less than $1\%$ between 5 and 30 bins. The concentration improves by $5\%$ between 5 and 15 bins and then flattens, with a further $0.5\%$ gain at 30 bins. The miscentering improves more steadily, by about $3\%$ at each step, so 30 bins are slightly better than 15 for this parameter alone. We keep 15 bins for the fiducial analysis. These results are shown in Fig.~\ref{fig:resolution_sigma}.

\section{Coverage tests}
\label{app:coverage}

\begin{figure*}[htb]
    \centering
    \includegraphics[width=\textwidth]{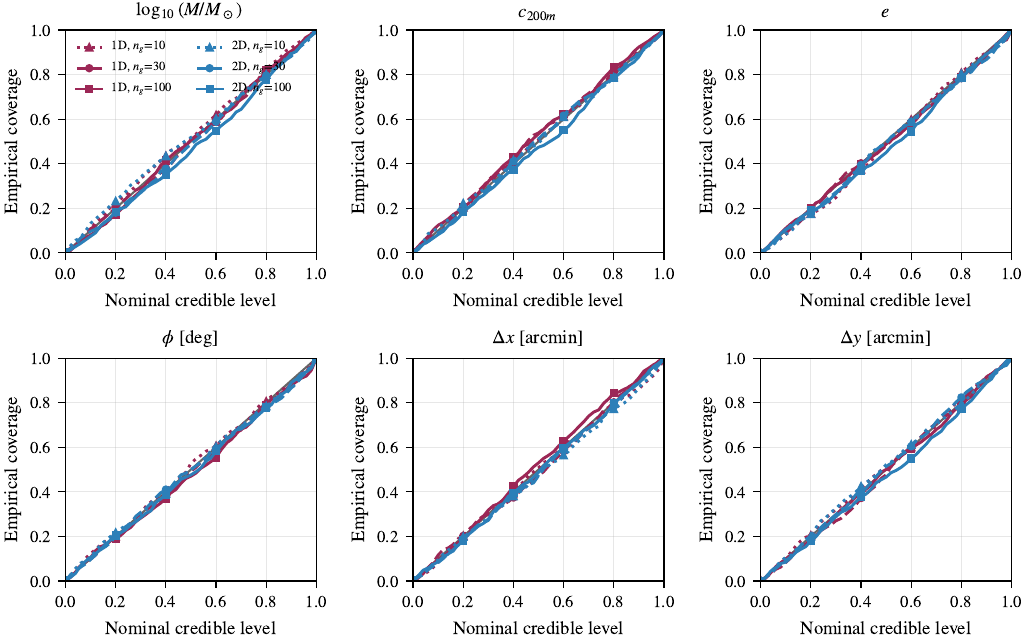}
    \caption{Empirical coverage against nominal credible level for each parameter, from 500 test simulations with $2000$ posterior samples each. Red: 1D summary; blue: 2D summary. Dotted, dashed and solid lines: $n_{\rm gal} = 10$, $30$ and $100\,{\rm arcmin}^{-2}$. Grey line: perfect calibration.}
    \label{fig:coverage}
\end{figure*}

We confirmed the accuracy of our learned posteriors using coverage tests. We create 500 test simulations within the prior and estimate the cluster posteriors for each simulation. We can then compare the empirical scatter to the estimated posteriors in order to assess whether the posteriors are under or over confident. These results are shown in Fig.~\ref{fig:coverage}. Each of the methods follow closely the diagonal, showing that the posteriors are correct.

\end{document}